# Anisotropic exchange coupling and ground state phase diagram of Kitaev compound YbOCl

Zheng Zhang,[1,2] Yanzhen Cai,[3] Jing Kang,[3] Zhongwen Ouyang,[4] Zhitao Zhang,[5] Anmin Zhang,[3] Jianting Ji,[1] Feng Jin,[1] and Qingming Zhang[3,1,*]

[1]*Beijing National Laboratory for Condensed Matter Physics, Institute of Physics, Chinese Academy of Sciences, Beijing 100190, China*
[2]*Department of Physics, Renmin University of China, Beijing 100872, China*
[3]*School of Physical Science and Technology, Lanzhou University, Lanzhou 730000, China*
[4]*Wuhan National High Magnetic Field Center and School of Physics, Huazhong University of Science and Technology, Wuhan 430074, China*
[5]*Anhui Province Key Laboratory of Condensed Matter Physics at Extreme Conditions, High Magnetic Field Laboratory,
Chinese Academy of Sciences, Hefei 230031, China*

Rare-earth chalcohalide REChX (RE = rare earth; Ch = O, S, Se, Te; X = F, Cl, Br, I) is a newly reported family of Kitaev spin liquid candidates. The family offers a platform where a strong spin-orbit coupling meets a van der Waals layered and undistorted honeycomb spin lattice, which outputs highly anisotropic exchange couplings required by the Kitaev model. YbOCl is the first single crystal of the family we grew, with a size up to ∼15 mm. We have performed magnetization and high magnetic field electron spin resonance measurements from 2 to 300 K. We develop the mean-field scenario for the anisotropic spin system, with which we are able to well describe the experiments and reliably determine the fundamental parameters. The self-consistent simulations give the anisotropic spin-exchange interactions of $J_\pm$ (∼ −0.3 K) and $J_{zz}$ (∼1.6 K), and $g$ factors of $g_{ab}$ (∼3.4) and $g_c$ (∼2.9). Based on the spin-exchange interactions, we employ the exact diagonalization method to work out the ground state phase diagram of YbOCl in terms of the off-diagonal exchange couplings. The phase diagram hosting rich magnetic phases including the spin-disordered one, sheds light on the novel magnetic properties of the family, particularly the Kitaev physics.

## I. INTRODUCTION

The exactly solvable Kitaev model defined on the honeycomb lattice gives novel quantum spin liquid (QSL) states [1] driven by anisotropic spin-exchange interactions rather than geometric frustration and has a promising potential in topological quantum computing. On the other hand, a material realization of the Kitaev model currently remains a significant challenge due to the unusual configuration of highly anisotropic bond-dependent spin interactions. Jackeli and Khaliullin proposed a scenario to produce the anisotropic spin interactions through strong spin-orbit coupling [2].

Several Kitaev QSL candidates based on the Jackeli-Khaliullin scenario have been explored. Among them, the 5$d$ iridate $A_2$IrO$_3$ ($A$ = Li, Na, and Cu) [3–8] is the first proposed family. However, many experiments have suggested that the ground state of this material family is a magnetically ordered state. The 4$d$ compound $\alpha$-RuCl$_3$ [9–12] has attracted extensive research interest. Although the ground state of the material is also a magnetically ordered state, a number of experiments have shown that magnetic fields may push $\alpha$-RuCl$_3$ into a spin-disordered state, which is considered to likely be a Kitaev QSL state.

Inspired by our previous studies of rare-earth-based spin frustrated materials, including the triangular lattice QSL candidates of rare-earth chalcogenides [13–16] and YbMgGaO$_4$ [17–20], we extend our material exploration for Kitaev QSL to rare-earth magnetic systems. This leads us to the discovery of a family of Kitaev QSL candidates, namely, rare-earth chalcohalides REChX (RE = rare earth; Ch = O, S, Se, and Te; X = F, Cl, Br, and I) [21]. Compared to the previously studied transition-metal Kitaev QSL candidates, rare-earth chalcohalides have many advantages. First, the building blocks are rare-earth magnetic ions with much stronger spin-orbit coupling (SOC) and more significant magnetic anisotropy [22,23]. Strong magnetic anisotropy is one of the basic requirements to achieve Kitaev QSL. Second, rare-earth ions with an odd number of 4$f$ electrons are protected by time-reversal symmetry, which gives rise to the formation of Kramers doublets. The Kramers doublets protected by time-reversal symmetry are robust and less affected by lattice distortions or other structural defects. Third, the honeycomb layers formed by magnetic ions in rare-earth chalcohalides are van der Waals layered and thus exhibit a true two dimensionality. This means that the study of atomically thin films of the family is possible, just like graphene. DyOCl, as a representative of this family of materials, has a van der Waals layered tetragonal structure. The single-layer or few-layer flakes of DyOCl were successfully prepared by mechanical exfoliation [24]. Finally, the honeycomb lattice formed by magnetic ions in rare-earth chalcohalides is an undistorted hexagonal structure [21] that meets the symmetry required by the *J-K-Γ* spin Hamiltonian model [25]. These features render rare-earth chalcohalide RECh*X* as a promising candidate family for the exploration of Kitaev QSL.

---

*qmzhang@ruc.edu.cn

In this paper, we focus on magnetism at finite temperatures and the ground state phase diagram of YbOCl. We have successfully synthesized high-quality YbOCl single crystals through a standard flux method [21]. We perform careful magnetization measurements for YbOCl single crystals. By applying the mean-field scenario to the low-energy effective spin Hamiltonian, we can simulate the magnetization results and determine the spin-exchange interactions of $J_\pm$ and $J_{zz}$, and the anisotropic Landé factors. The simulations are cross checked with a 12-site full diagonalization (FD) method.

We have carried out the measurements of high magnetic field electron spin resonance (ESR). The Landé factors given by the ESR measurement are in good agreement with the results obtained from the above magnetization measurements. Interestingly, the *g* factors read from ESR spectra show a clear temperature dependence, which arises from the fact that the exchange couplings gradually play a role in decreasing temperatures. By generalizing the mean-field scenario to the ESR measurements, we have well understood the temperature dependence of the anisotropic Landé factors. This allows us to develop an alternative way to check the reliability of the fundamental spin-exchange parameters and *g* factors.

The studies set a reliable experimental basis for our exploration of the theoretical ground state phase diagram of YbOCl. By combining the spin-exchange interactions of $J_\pm$ and $J_{zz}$ with a 24-site exact diagonalization (ED) method, we work out the ground state phase diagram of YbOCl in terms of the off-diagonal exchange parameters of $J_{\pm\pm}$ and $J_{z\pm}$. The rich phase diagram contains not only various magnetically ordered phases, but also a spin-disordered one. We further investigate the effect of Dzyaloshinskii-Moriya (DM) interaction on the ground state phases. The next-nearest-neighbor (NNN) DM interaction has little influence on the ground state phase diagram if it is less than 0.4 K. The 120-degree region will be expanded under a higher DM interaction.

## II. SAMPLES, EXPERIMENTAL TECHNIQUES, AND CALCULATION METHODS

The high-quality single crystals of YbOCl have been grown by the anhydrous YbCl$_3$-flux method [21]. The typical size of the single crystals is about $6 \times 8 \times 0.05$ mm$^3$ [26], with a maximum size up to 15 mm.

We performed x-ray diffraction (XRD) and energy dispersive x-ray (EDX) measurements on polycrystalline and single-crystal samples of YbOCl, respectively, to confirm the quality of YbOCl. The Rietveld refinement residual $R_{wp}$ for the XRD pattern of the polycrystalline sample was 9.06%. The EDX measurement result shows that the ratio Yb:O:Cl is 0.9:1:1 (see Supplemental Material [26]).

Approximately 4 mg of YbOCl single crystals have been prepared to perform M/H-T and magnetization measurements. The anisotropic measurements along the *c* axis and in the *ab* plane of YbOCl have been performed using a Quantum Design physical property measurement system (PPMS) from 1.8 to 300 K under magnetic fields of 0 to 7 T. Approximately 3 mg of single crystals have been used to perform pulsed high magnetic field ESR measurements. The pulsed magnetic fields are generated by a coil driven by a 90 kJ capacitor bank. The ESR data are collected using a spectrometer with operating frequencies of ∼65.4-300 GHz and at a temperature range of ∼2-20 K. The magnetic fields are applied in the *ab* plane and along the *c* axis. Approximately 3 mg of single crystals have been used in the conventional ESR measurements, which are performed with a Bruker EMX plus 10/12 continuous-wave spectrometer at *X*-band frequencies ($f \sim 9.39$ GHz).

We develop the mean-field (MF) theory for the anisotropic spin Hamiltonian and derive the self-consistent equations to simulate M/H-T, magnetization, and ESR results. For details related to the MF method, please refer to the Supplemental Material [26]. The 12-site FD method is employed to simulate M/H-T data. We use the 24-site ED method to calculate the ground state phase diagram of YbOCl. For details related to the FD and ED methods, please also refer to the Supplemental Material [26].

## III. CRYSTALLINE ELECTRIC FIELD ENVIRONMENTS AND ANISOTROPIC SPIN HAMILTONIAN

The rare-earth magnetic ion Yb$^{3+}$ in YbOCl possesses an electronic configuration of $4f^{13}$, which gives a spectral terms of $^2F_{7/2}$ due to the strong spin-orbit coupling of $4f$ electrons [16]. In YbOCl, three Cl$^-$ and four O$^{2-}$ around Yb$^{3+}$ form a crystalline electric field (CEF) environment with a point group symmetry of $C_{3v}$. To estimate the temperature range in which Yb$^{3+}$ can be taken as an effective spin-1/2, we carry out the CEF excitation energy levels in YbOCl based on the point charge model (PCM). The calculations show that the energy separation between the CEF ground state and the first excited state is more than 60 meV [26]. We can also see that the single crystal of YbOCl is transparent [26], which indicates that YbOCl is a typical ionic crystal. Generally, ionic crystals have higher CEF excitation energy levels, indirectly supporting that YbOCl has high CEF excitation energy levels. This ensures that the unpaired electrons in YbOCl can be safely regraded as an effective spin-1/2 at least below room temperature.

To investigate the magnetism in YbOCl, which is dominated by its low-energy spin excitations, we start from the low-energy effective spin Hamiltonian after carefully ruling out the effect of the CEF excitations as discussed above. The crystal structure and the symmetry of YbOCl basically determine the form of its low-energy effective spin Hamiltonian. The honeycomb magnetic layer in YbOCl is formed by two closely adjacent triangular layers. The superexchange interaction between the nearest-neighbor (NN) magnetic ions is mediated by O$^{2-}$ anions and there is no first-order Dzyaloshinskii-Moriya (DM) interaction because of the existence of an inversion center between the NN magnetic ions. The absence of an inversion center between the next-nearest-neighbor (NNN) magnetic ions allows the second-order DM interaction. We have determined that the DM vector $\vec{D}_{ij}$ is almost along the *c* axis [25]. The analysis gives the following low-energy effective spin Hamiltonian in

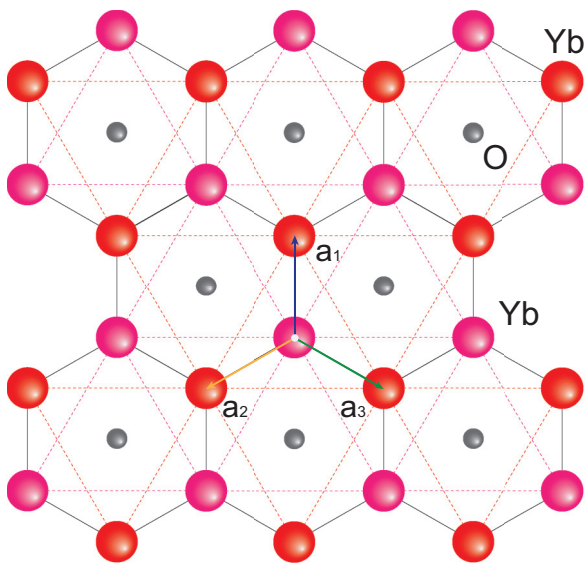

FIG. 1. The honeycomb lattice formed by Yb$^{3+}$ in YbOCl. The three directions of $a_1$, $a_2$, and $a_3$ are related to the phase factors defined on the honeycomb lattice in the low-energy effective spin Hamiltonian.

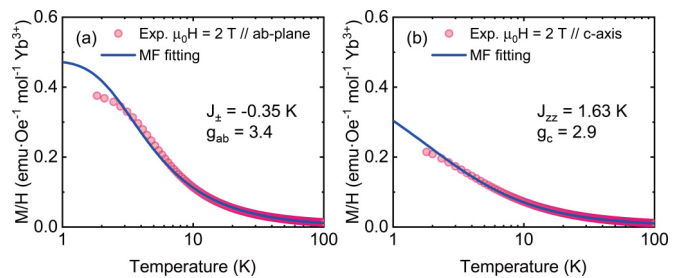

FIG. 2. Temperature dependence of M/H (a) in the $ab$ plane and (b) along the $c$ axis in YbOCl and mean-field (MF) fitting. The red circles are experimental results and the blue and green solid lines are MF fitting results. $J_\pm$, $J_{zz}$, $g_{ab}$, and $g_c$ shown here are the fitting results optimized with the nonlinear least-square method. The deviation between experiments and calculations becomes bigger at low temperatures, which indicates that the off-diagonal spin interactions come into play.

YbOCl:

$$\begin{aligned}
\hat{H} = & \sum_{\langle ij \rangle} \big[ J_{zz} S_i^z S_j^z + J_\pm (S_i^+ S_j^- + S_i^- S_j^+) \\
& + J_{\pm\pm}(\gamma_{ij} S_i^+ S_j^+ + \gamma_{ij}^* S_i^- S_j^-) \\
& + J_{z\pm}\big(\gamma_{ij} S_i^+ S_j^z + \gamma_{ij}^* S_i^- S_j^z + \langle i \longleftrightarrow j \rangle \big) \big] \\
& + \sum_{\langle\langle ij \rangle\rangle} \vec{D}_{ij} \cdot (\vec{S}_i \times \vec{S}_j) \\
& - \mu_0 \mu_B \sum_i \big[ g_{ab}(h_x S_i^x + h_y S_i^y) + g_c h_c S_i^z \big],
\end{aligned} \quad (1)$$

where $J_{zz}$, $J_\pm$, $J_{\pm\pm}$, and $J_{z\pm}$ are the NN anisotropic spin-exchange parameters, $\vec{D}_{ij}$ is the NNN DM interaction, $g_{ab}$ and $g_c$ represent the Landé factors in the $ab$ plane and along the $c$ axis, and $\gamma_{ij}$ is 1, $e^{2i\pi/3}$, and $e^{-2i\pi/3}$ on the bonds along the $\vec{a}_1$, $\vec{a}_2$, and $\vec{a}_3$ directions, respectively, as shown in Fig. 1. The low-energy effective spin Hamiltonian is the starting point to analyze the magnetism of YbOCl. By applying appropriate numerical or approximation methods to the Hamiltonian, we are able to not only describe the thermodynamics at finite temperatures, but also explore the ground state phase diagram. It is worth noting that the low-energy effective spin Hamiltonian adopted in this paper is a generic one and completely equivalent to the $J$-$K$-$\Gamma$ model. The two Hamiltonians can be converted to each other through coordinate transformation [25,27]. In this paper, we choose the generic low-energy effective spin Hamiltonian described in formula (1), as $S_z$ in the formula is consistent with the application of magnetic fields along the $c$ axis of single crystals. Formula (1) is more convenient for us to analyze the experimental results.

## IV. MAGNETIZATIONS AT FINITE TEMPERATURES

As mentioned above, the PCM calculations tell us that the first CEF excitation energy is greater than 60 meV. Such a large CEF energy gap ensures that the basic picture of effective spin-1/2 is valid at least below 300 K. Then, one can estimate the spin-exchange interactions by fitting the M/H-T data at finite temperatures. To fit and calculate the magnetization of YbOCl, we apply MF approximation to the low-energy spin Hamiltonian, which is described in formula (1) [26]. It should be noted that the off-diagonal spin-exchange interactions $J_{\pm\pm}$ and $J_{z\pm}$ in formula (1) are actually canceled out at the MF level and can be retrieved using a quantum numerical method such as quantum Monte Carlo or ED. We further neglect the DM interaction in formula (1), considering that the DM interaction is just the NNN term. So far, we can derive the MF form of the spin Hamiltonian with magnetic fields in the $ab$ plane and along the $c$ axis, respectively [26]. Here the anisotropic Landé $g$ factors of $g_{ab}$ in the $ab$ plane and $g_c$ along the $c$ axis are also undetermined parameters and should be taken together with the spin-exchange parameters $J_\pm$ and $J_{zz}$ to make an optimal fitting. The experimental magnetizations and the fitting curves, both in the $ab$ plane and along the $c$ axis, are shown in Fig. 2. For magnetic fields in the $ab$ plane and along the $c$ axis, the optimal fitting gives $J_\pm$ of $-0.35$ K and $g_{ab}$ of 3.4, $J_{zz}$ of 1.63 K and $g_c$ of 2.9, respectively. According to the experimental data and the fitting results, we also estimate the fitting errors in the $ab$ plane and along the $c$ axis. The fitting error for the spin-exchange interaction $J_\pm$ and the Landé factor $g_{ab}$ is $\sim 1.98\%$. The fitting error for the spin-exchange interaction $J_{zz}$ and the Landé factor $g_c$ is $\sim 0.50\%$.

We further measure the field dependence of magnetizations at 10, 20, and 50 K. The M-H results can be well reproduced with the MF approximation using the spin-exchange parameters and $g$ factors obtained above. The measured M-H data and the calculated M-H curves are shown in Fig. 3. The MF results are consistent with the experimental ones. There is a slight deviation between the experiments and the calculations at 10 K under high magnetic fields, which may be related to the MF approximation ignoring the spin-exchange parameters $J_{\pm\pm}$, $J_{z\pm}$, and the DM interaction.

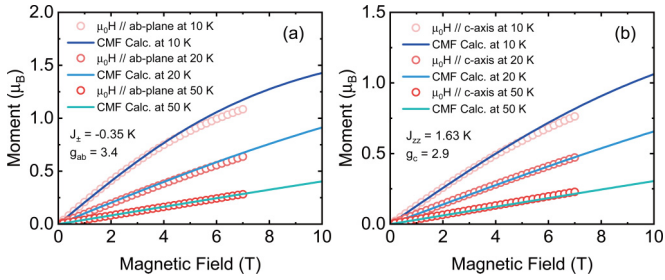

FIG. 3. (a) M-H curves in the *ab* plane at selected temperatures. (b) M-H curves along the *c* axis at selected temperatures. The open circles are experimental data and the solid lines are the MF simulations. Note that the larger deviation between the experiment data and MF simulations at lower temperatures originates from the off-diagonal spin interactions. See the text for details.

The spin-exchange parameters and *g* factors obtained by the above MF fitting have been cross checked by the FD method that takes the spin-correlation effect. Currently, the method is limited to spin systems with a small number of spins. Here we treat a honeycomb spin lattice composed of 12 spins with periodic boundary conditions [see Fig. 4(a)]. In the FD fitting, the contribution of the anisotropic spin-exchange interaction $J_{\pm\pm}$, $J_{z\pm}$, and DM interaction are ignored, and the Landé factors $g_{ab} = 3.4$ and $g_c = 2.9$ from the MF fitting are adopted.

The FD fitting demonstrates that the absolute values of the spin-exchange interactions $J_{\pm}$ and $J_{zz}$ do not exceed 2 K. Therefore, the fitting interval of the two spin-exchange interactions can be safely limited to the range of −2 to 2 K. The errors given by the FD fitting are shown in Figs. 4(d) and 4(e). The formula for estimating normalized errors can be written as

$$R = N\left[\frac{|M/H_{exp} - M/H_{cal}|}{M/H_{exp}}\right]. \quad (2)$$

The white dashed lines in Figs. 4(f) and 4(g) trace the optimal values of $J_{\pm}$ and $J_{zz}$. The combination of the two lines finally determines $J_{\pm}$ and $J_{zz}$ to be −0.3 and 1.6 K, respectively. The values are very close to the results obtained by the MF method. We plot the FD calculation results in Fig. 4(b)–4(e). It can be seen that the calculations are basically consistent with the experimental results, especially the M/H-T data with a magnetic field of 2 T. This suggests that without the off-diagonal spin-exchange parameters, the spin-exchange interactions of $J_{\pm}$ and $J_{zz}$ are sufficiently good to describe the general thermodynamics at finite temperatures, such as the temperature and field evolutions of magnetizations. More importantly, $J_{\pm}$ and $J_{zz}$ given by the MF fitting and FD fitting show little difference, strengthening the reliability of the spin-exchange parameters obtained here. Compared with the MF method, the FD method seems to bring an insignificant improvement, indicating that spin correlation may play a minor role in the magnetism at finite temperatures and under high magnetic fields.

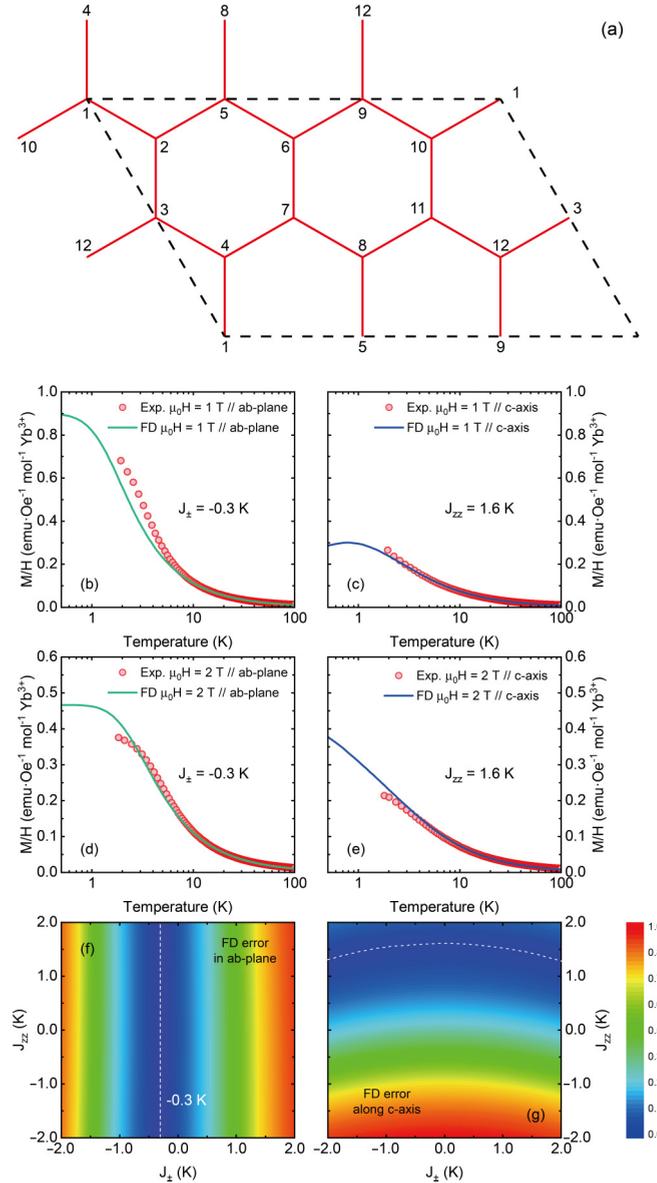

FIG. 4. (a) Honeycomb lattice with 12 spins and periodic boundary conditions. (b),(c) Temperature dependence of M/H with a magnetic field in the *ab* plane. (d),(e) Temperature dependence of M/H with a magnetic field along the *c* axis. The open circles are experimental data and the solid lines are the FD simulations. As seen in Figs. 2 and 3, the deviation between experiments and simulations also increases with lowering temperatures. This stems from the effect of off-diagonal spin interactions and is discussed in the text. (f),(g) The error *R* between the experiments and the FD results in the *ab* plane and along the *c* axis under a magnetic field of 1 T. *R* is defined by Eq. (2) in the context. The white dashed lines trace the optimal values. (f) and (g) share the same color scale in the lower right corner.

## V. HIGH MAGNETIC FIELD ELECTRON SPIN RESONANCE

The anisotropic *g* factors can be directly read out of the ESR spectra, containing crucial information on the spin interactions. In this sense, we can construct a complete and self-consistent scenario to reliably determine the fundamental parameters and study the magnetism at finite temperatures [28,29] by combining thermodynamics and spectral (ESR) measurements.

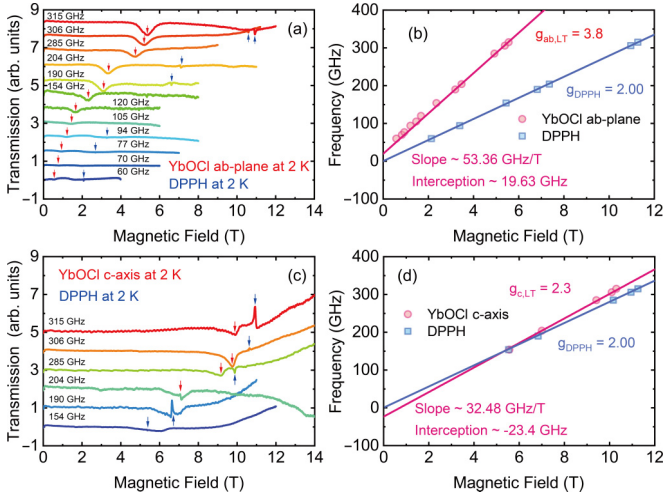

FIG. 5. (a),(c) Frequency-dependent ESR spectra measured at 2 K in the *ab* plane and along the *c* axis. The resonance signals from YbOCl single crystals are marked by the red arrows, while the blue arrows mark the resonance signals from the reference sample. The 1, 1-diphenyl-2-picrylhydrazyl (DPPH), which has a standard *g* factor of 2.00 for calibration. (b),(d) Field dependence of resonance frequencies. The red and blue circles refer to YbOCl and DPPH, respectively. Correspondingly, the red and blue solid lines are the linear fitting results for them. Considering that the ESR signals at low frequencies are relatively weak, for accuracy we carefully choose the data at high frequencies ($\geqslant 190$ GHz) for linear fitting.

We first perform the conventional X-band ESR measurements on the YbOCl single crystals. Unfortunately, the ESR signal is incomplete on an ESR spectrometer at 9.4 GHz [26] since the ESR linewidth exceeds 1 T. Then, we perform pulsed high magnetic field ESR measurements with magnetic fields up to 14 T.

The ESR spectra at 2 K in the *ab* plane and along the *c* axis are shown in Figs. 5(a) and 5(c), respectively. It can be seen that the intensities of the ESR signals generally increase with increasing frequencies, and the ESR signals in the *ab* plane are significantly stronger than those along the *c* axis, indicating a strong magnetic anisotropy in YbOCl. One can extract the field dependence of resonance frequencies as shown in Figs. 5(b) and 5(d). The linear fitting results give the slope of 53.36 GHz/T and the interception of 19.63 GHz in the *ab* plane; and the slope of 32.48 GHz/T and the interception of $-23.40$ GHz along the *c* axis. Using these fitting parameters and the following formula, one can calculate $g_{LT}$ at low temperatures:

$$g_{LT} = \frac{h\nu}{\mu_B B}, \qquad (3)$$

where $h$ is the Planck constant, $\nu$ the resonance frequency, and $B$ the resonance magnetic field. It should be noted that the calculated anisotropic $g$ factors are 3.8 in the *ab* plane and 2.3 along the *c* axis at 2 K, which deviate from the values at higher temperatures. This stems from the fact that spin-exchange interactions play a dominant role at low temperatures, and hence the low-temperature *g* factors contain crucial information on the spin interactions. In other words, at low temperatures, the resonance field *B* is not only contributed by an external field, but also by an internal field induced by spin-exchange interactions. We can quantitatively understand this at a MF level. The low-energy effective spin Hamiltonian described by formula (1) can be written under the MF approximation as follows:

$$\hat{H}_{ab} = \left(\frac{6J_{\pm}M_{ab}}{\mu_0 \mu_B g_{ab}} - \mu_B g_{ab} B_{ab}\right) \sum_i S_i^x - \frac{3J_{\pm}N(M_{ab})^2}{(\mu_0 \mu_B g_{ab})^2},$$

$$\hat{H}_c = \left(\frac{3J_{zz}M_c}{\mu_B g_c} - \mu_B g_c B_c\right) \sum_i S_i^z - \frac{3J_{zz}N(m_z)^2}{2(\mu_0 \mu_B g_c)^2}, \qquad (4)$$

where $M_{ab}$ and $M_c$ are magnetic moments in the *ab* plane and along the *c* axis, respectively. The magnetic moments are given by the M/H-T measurements or the calculations by self-consistent equations. We can get an effective field by extracting the common factors $-\mu_B g_{ab}$ (or $-\mu_B g_c$) from formula (4),

$$B_{ab,eff} = B_{ab} - \frac{6J_{\pm}M_{ab}}{\mu_B^2 g_{ab}^2},$$

$$B_{c,eff} = B_c - \frac{3J_{zz}M_c}{\mu_B^2 g_c^2}. \qquad (5)$$

Substituting the effective field into formula (3), now we have a more accurate relationship between the resonance frequency $\nu$ and external magnetic field $B_{ab}$ (or $B_c$),

$$\nu = \frac{\mu_B g_{ab,LT}}{h} B_{ab} - \frac{6J_{\pm}M_{ab}g_{ab,LT}}{\mu_B g_{ab}^2 h},$$

$$\nu = \frac{g_{c,LT}\mu_B}{h} B_c - \frac{3J_{zz}M_c g_{c,LT}}{\mu_B g_c^2 h}. \qquad (6)$$

Using the slopes given by the linear fitting in Figs. 5(b) and 5(d), we can directly calculate $g_{ab,LT} \sim 3.8$ (an error of 0.41%) and $g_{c,LT} \sim 2.3$ (an error of 0.96%), which are well consistent with formula (3). More importantly, the interceptions in formula (6) are caused by spin-exchange interactions $J_{\pm}(\sim -0.35$ K) and $J_{zz}(\sim 1.63$ K), the Landé factors $g_{ab}$ ($\sim 3.4$) and $g_c$ ($\sim 2.9$) of the isolated magnetic ion, and magnetic moments $M_{ab}$ ($\sim 7500/N_A$ emu) and $M_c$ ($\sim 4420/N_A$ emu), which have been given by fitting the M/H-T data or magnetization data. Independent of the slopes, the interceptions also give $g_{ab,LT} \sim 3.9$ and $g_{c,LT} \sim 2.4$. The perfect consistency suggests that thermodynamics measurements plus ESR spectra offer a self-consistent scenario to describe the magnetism at finite temperatures at the MF level, which can be generalized to other anisotropic spin systems.

We further carry out temperature-dependent ESR measurements with fixed frequencies of 315 GHz in the *ab* plane and 306 GHz along the *c* axis.

The resonance spectra in the *ab* plane and along the *c* axis are shown in Figs. 6(a) and 6(c), respectively. We can well simulate the temperature dependence of the $g_{LT}$ at the MF level. We can derive the following formula from the above

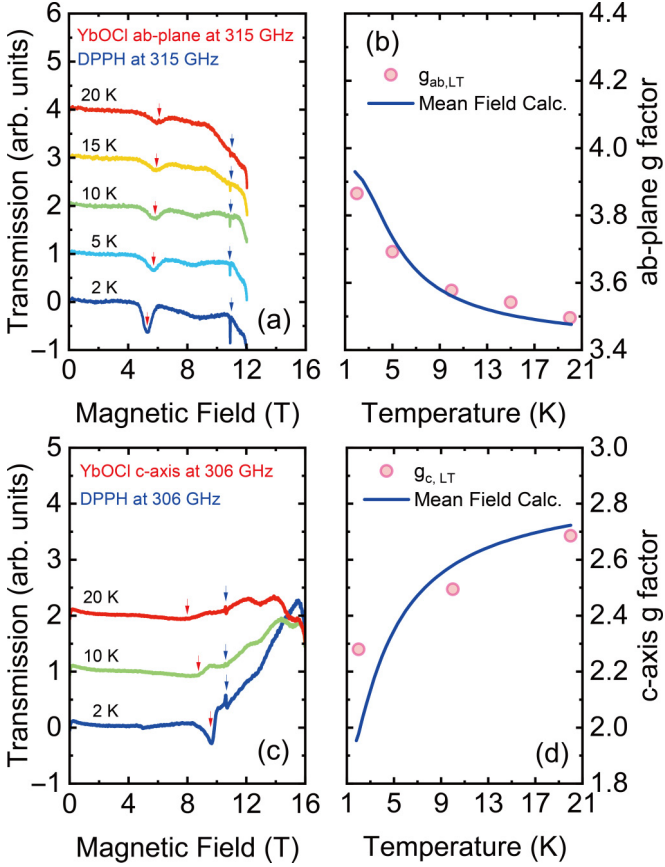

FIG. 6. (a), (c) Temperature-dependent ESR spectra measured at 315 GHz in the *ab* plane and 306 GHz along the *c* axis. The red arrows mark the resonance fields in YbOCl and the blue arrows mark the resonance fields of the reference sample DPPH. (b),(d) The red circles and the blue lines are the $g_{exp}$ data obtained at different temperatures and the MF calculation results, respectively.

MF expressions:

$$g_{ab,LT} = \frac{h\nu}{\mu_B B_{ab,eff}} = \frac{h\nu}{\mu_B \left(B_{ab} - \frac{6J_\pm M_{ab}}{\mu_B^2 g_{ab}^2}\right)},$$

$$g_{c,LT} = \frac{h\nu}{\mu_B B_{c,eff}} = \frac{h\nu}{\mu_B \left(B_c - \frac{3J_{zz} M_c}{\mu_B^2 g_c^2}\right)}. \quad (7)$$

With the above formula, we can estimate the internal fields contributed by magnetic moments and spin-exchange interactions. The internal fields in the *ab* plane and along the *c* axis are 0.36 and −0.7 T, respectively. The highest applied or external resonance magnetic fields in the *ab* plane and along the *c* axis are about 5.5 and 11.2 T, respectively. Although the external magnetic fields are higher than the internal fields, we cannot ignore the effect of the internal field on the ESR peaks and $g$ factors, particularly at low temperatures. By extracting the common factor $-B\mu_B$ from the MF Hamiltonian described in formula (4), we find an accurate relationship between the low-temperature $g_{LT}$ factors and the ones in the case of isolated magnetic moments or high temperatures, just as follows:

$$g_{ab,LT} = g_{ab} - \frac{6J_\pm M_{ab}}{\mu_B^2 g_{ab} B_{ab}},$$

$$g_{c,LT} = g_c - \frac{3J_{zz} M_c}{\mu_B^2 g_c B_c}. \quad (8)$$

Clearly, the temperature-dependent $g_{LT}$ factors approach the isolated-moment ones with temperatures, as magnetizations become negligibly small with increasing temperatures. The calculated $g_{LT}$ factors are shown in Figs. 6(b) and 6(d) using the above formula, which agrees with the experimental results. The slight differences between the calculated and experimental results may be related to several reasons. (1) The ESR line width of YbOCl is quite wide, and the resonance signals decay rapidly as the temperature increases. This may cause some errors in reading the resonance peak positions. (2) $g_{LT}$ is related to the spin-exchange interactions at low temperatures. The MF approximation cancels out the off-diagonal spin interactions of $J_{\pm\pm}$ and $J_{z\pm}$. This may also bring some errors. Regardless, formula (8) based on the MF method well explains the behavior of $g_{LT}$ with varying temperatures.

It should be pointed out that the combination of thermodynamics and ESR measurements actually gives redundant but independent determinations for the anisotropic exchange parameters of $J_\pm$ and $J_{z\pm}$ and $g$ factors. This is not only a robust guarantee for the reliability of the parameters, but also can be used to accurately describe the magnetism of anisotropic spin systems at finite temperatures. Furthermore, the high magnetic field ESR spectra confirm that the NNN spin-exchange interactions are weak in YbOCl. The simple reason is that more ESR peaks in the *ab* plane and along the *c* axis should be observed if the NNN spin-exchange interactions are comparable to the NN spin-exchange interactions.

## VI. GROUND STATE PHASE DIAGRAM WITHOUT DM INTERACTION

The above study allows us to explore the ground state phase diagram by employing the ED technique. By fixing $J_\pm = -0.3$ K and $J_{zz} = 1.6$ K, we first work out the phase diagram without DM interaction in terms of the off-diagonal spin interactions $J_{\pm\pm}$ and $J_{z\pm}$. In the ED calculations, we take a honeycomb lattice with 24 spin sites and periodic boundary conditions (PBCs). The honeycomb lattice is shown in Fig. 7(a). To accelerate the ED calculations, we adopt a single tensor to represent the ground state wave function and a $2 \times 2 \times 2 \times 2$ tensor to describe the local two-body spin operators in the low-energy effective spin Hamiltonian. By doing so, we need to store only the ground state wave function rather than the $2^{24} \times 2^{24}$ matrices representing the spin Hamiltonian in memory. This greatly saves memory and is particularly beneficial for parallel computing. The details of the calculations can be found in the Supplemental Material [26].

The spin-exchange parameters $J_{\pm\pm}$ and $J_{z\pm}$ are limited to −2 to 2 K. The ED calculations in the parameter space reveal a rich ground state phase diagram that contains seven distinct magnetic phases (Fig. 7). The real-space spin patterns for the magnetic phases are illustrated at the bottom of Fig. 7. Corresponding to the real-space patterns, the static spin structure

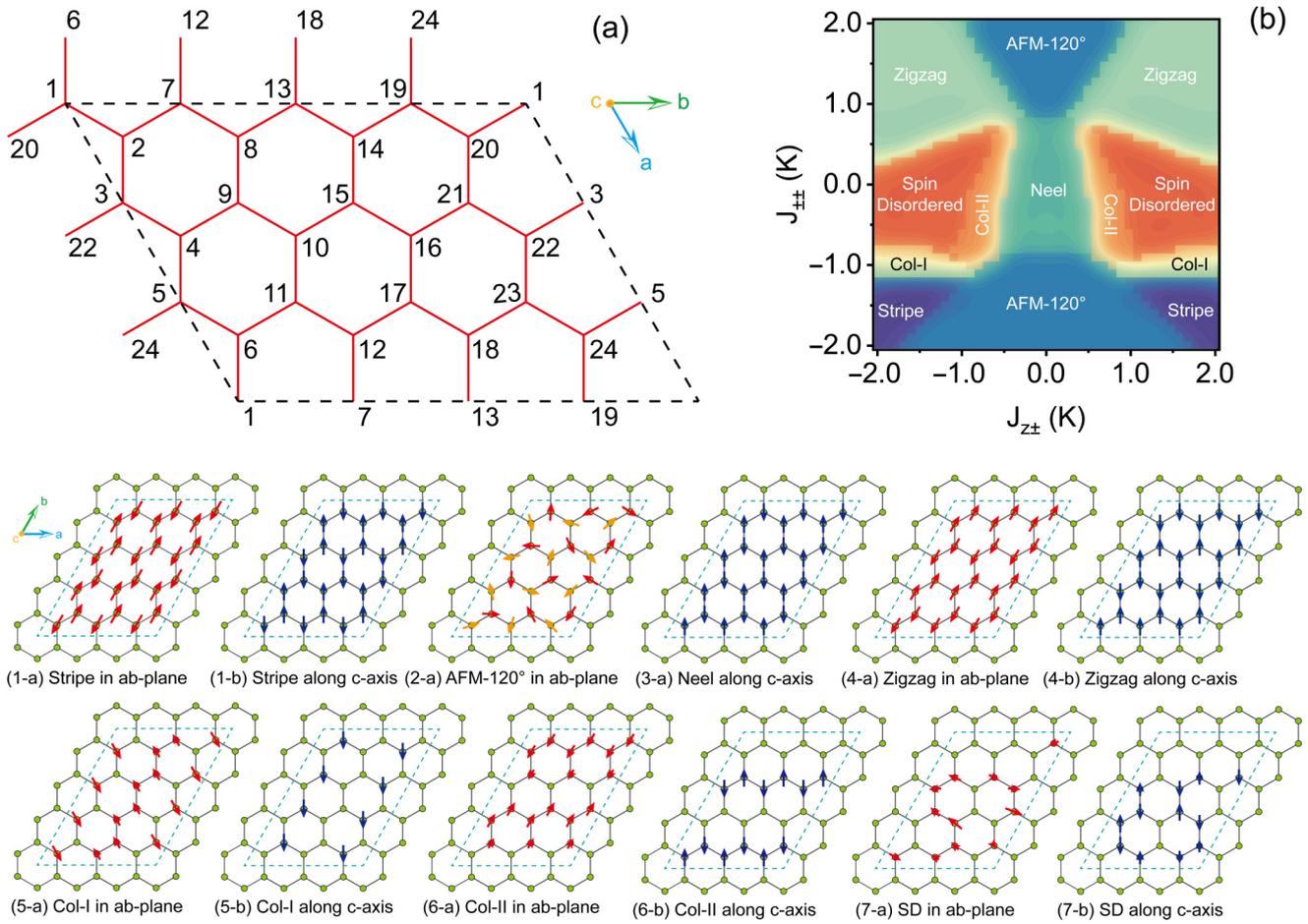

FIG. 7. (a) Honeycomb lattice with 24 spin sites and periodic boundary conditions taken in our ED calculations for the ground state phase diagram. (b) Ground state phase diagram of YbOCl. The real-space spin patterns of the magnetic phases in the phase diagram are illustrated in the following rows. AFM, Col-I, Col-II, and SD refer to antiferromagnetic, columnar-I, columnar-II, and spin-disordered phases, respectively.

factors for the magnetic phases in the reciprocal space are further calculated and shown in Fig. 8. The static structure factor is defined as

$$\chi_N^\nu(\vec{Q}) = \frac{1}{N} \sum_{ij} e^{i\vec{Q}\cdot(\vec{R}_i - \vec{R}_j)} \langle S_i^\nu S_j^\nu \rangle, \quad (9)$$

where the spin component $\nu = x, y, z$, $R_i$ is the coordinate of lattice site $i$, $N$ is the total number of lattice sites, and $\langle S_i^\nu S_j^\nu \rangle$ is the spin correlation between site $i$ and site $j$. In the following, we will examine and discuss the magnetic phases in the phase diagram one by one in terms of the real-space spin

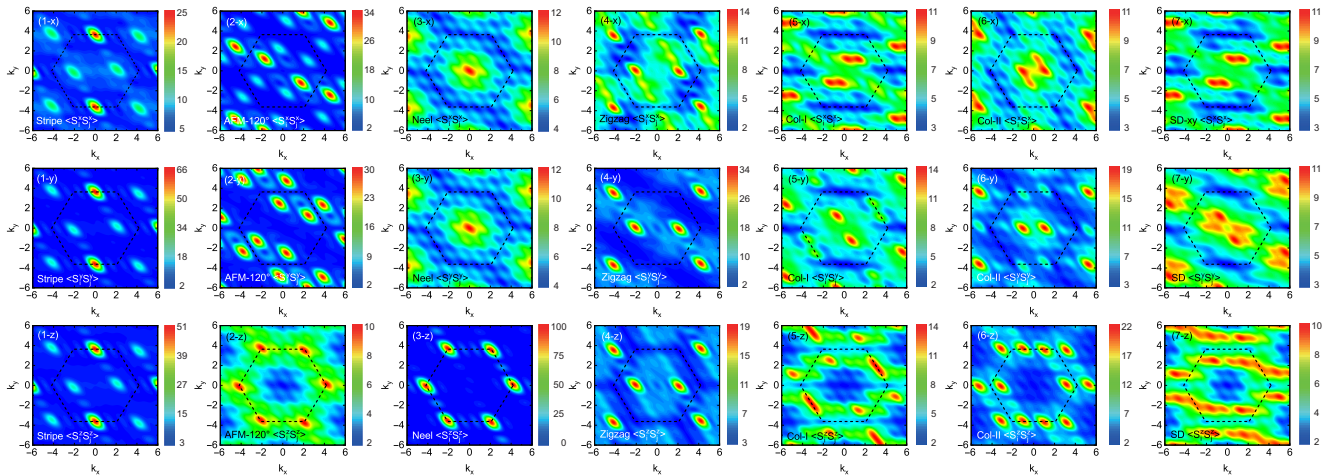

FIG. 8. The static structure factors corresponding to the seven magnetic phases in the ground state phase diagram.

patterns plus the static spin structure factors in the reciprocal space.

(1) Stripe phase. In this phase, the stripe arrangement can be seen not only in the *ab* plane [Fig. 7(1-a)], but also along the *c* axis [Fig. 7(1-b)]. Correspondingly, the three spin structure factors $\chi_N^x$, $\chi_N^y$, and $\chi_N^z$ are centered at the high-symmetry point $M$ in the reciprocal space, as shown in Figs. 8(1-x), 8(1-y), and 8(1-z).

(2) AFM-120° phase. The magnetic moments almost lie in the *ab* plane and little along the *c* axis. Generally, a honeycomb lattice can be regarded as a combination of two triangular sublattices. On each triangular sublattice, the angle between magnetic moments is about 120°. The magnetic moments on the two triangular lattices are marked in orange and red in Fig. 7(2-a). The intensities of the static spin structure factors $\chi_N^x$ [Fig. 8(2-x)] and $\chi_N^y$ [Fig. 8(2-y)] are about three times larger than those of $\chi_N^z$ [Fig. 8(2-z)], which again indicates that the magnetic moments in the phase lie in the *ab* plane.

(3) Neel phase. The dominant components of the magnetic moments in the phase are antiferromagnetically arranged along the *c* axis. The spin structure factor $S_N^z$ shows a strong excitation at the high-symmetry point $K$ [Fig. 8(3-z)]. Since there is a small portion of components in the *ab* plane, a weak excitation near the Γ point in the channels of $\chi_N^x$ [Fig. 8(3-x)] and $\chi_N^y$ [Fig. 8(3-y)] can be seen. The density-matrix renormalization group (DMRG) method has been applied to the Heisenberg-Γ model defined on the honeycomb lattice, which also gives the Neel state in the parameter range similar to what we adopt [30]. Interestingly, the DMRG studies indicate that the ground state for the *XXZ* model defined on the honeycomb lattice is also a Neel state [31]. For example, rare-earth magnetic material YbCl$_3$ with a honeycomb lattice has a typical Neel state [32–34]. This shows that the anisotropic spin-exchange interactions play a decisive role in the realization of a spin-disordered state or QSL in the honeycomb lattice having no geometrical spin frustration.

(4) Zigzag phase. The magnetic moments are arranged in a zigzag order both in the *ab* plane [Fig. 7(4-a)] and along the *c* axis [Fig. 7(4-b)]. And the excitation patterns revealed by $\chi_N^x$, $\chi_N^y$, and $\chi_N^z$ are similar, which correspond to a perfect zigzag order in the real space.

(5) Columnar-I (Col-I) phase. In the *ab* plane [Fig. 7(5-a)], the magnetic moments form a columnar ferromagnetic arrangement along the *b* direction, while it shows a complicated antiferromagnetic arrangement along the *a* direction. The components along the *c* axis also form a columnar ferromagnetic arrangement [Fig. 7(5-b)]. The spin structure factors $\chi_N^x$ [Fig. 8(5-x)], $\chi_N^y$ [Fig. 8(5-y)], and $\chi_N^z$ [Fig. 8(5-z)] also behave in a complicated way. The characteristics of continuous excitations can be observed at both high-symmetry points and non-high-symmetry points.

(6) Columnar-II (Col-II) phase. The Col-II phase looks like a version of Col-I in which the *a* direction and *b* direction are exchanged. In other words, the magnetic moments form a columnar ferromagnetic arrangement along the *x* direction and a complicated antiferromagnetic order along the *y* direction. The components along the *c* axis form a columnar antiferromagnetic arrangement [Fig. 7(6-b)]. Its spin structure factors $\chi_N^x$ [Fig. 8(6-x)], $\chi_N^y$ [Fig. 8(6-y)], and $\chi_N^z$ are similar to those of Col-I phase [Fig. 8(6-z)].

(7) Spin-disordered (SD) phase. There is no characteristic of a magnetic order both in the *ab* plane [Fig. 7(7-a)] and along the *c* axis [Fig. 7(7-b)]. Particularly for the spin structure factors $\chi_N^y$ [Fig. 8(7-y)] and $\chi_N^z$ [Fig. 8(7-z)], there exists a complex pattern. The spin-disordered phase is a crucial starting point to explore the quantum spin liquid in the anisotropic spin system.

There are more interesting features for the ground state phase diagram. First, it is symmetrical with respect to the $J_{z\pm}$ axis, which means that the sign of $J_{z\pm}$ has little effect on magnetic ordering. Second, the Col-I and Col-II phases are located at the edge of the SD phase, indicating that the two phases are transitional phases from the magnetically ordered states to the disordered one. Finally, anisotropic spin-exchange interactions play a key role in generating such a rich phase diagram containing a variety of magnetic structures.

In the phase diagram, the SD phase is of particular interest and deserves a more in-depth discussion. We investigate the evolution of spin correlation with distance. As shown in Fig. 9(a), the spin site No.1 is taken as the starting point and marked as the zero-order neighbor. The other spin sites are marked from the first-order neighbor to the eighth-order neighbor according to their distances from site No.1. We present more details in the Supplemental Material on classifying the spin sites by order [26]. A set of the zero-order neighbor to the eighth-order neighbor are spin sites of 1, 2, 3, 12, 4, 9, 17, 13, and 18, which are marked by colored circles in Fig. 6(a).

The stripe, Neel, and zigzag phases possess a short-range periodic structure. Their spin correlations also exhibits short-range oscillations [see Figs. 9(b), 9(d) and 9(e)]. The AFM-120°, Col-I, and Col-II phases have long-range or complex periodic structures, and the spin correlations also show no clear oscillations. For the SD phase, the spin correlation rapidly decays with increasing the distances (or the neighbor orders). The characteristic is similar to the rapid decay of the spin correlation with increasing distance obtained from the Kitaev model on small lattices [35]. It means that the disordered phase is related to short-range rather than long-range spin correlation.

## VII. GROUND STATE PHASE DIAGRAM WITH DM INTERACTION

Although the DM interaction $\vec{D}_{ij}$ in YbOCl is the NNN one, it generally favors the formation of a magnetically ordered state. Especially in the magnetic materials with 3*d* transition-metal magnetic ions, the valence changes of 3*d* ions tend to cause lattice distortion, which leads to DM interactions that cannot be ignored. For example, in the famous kagome compound ZnCu$_3$(OH)$_6$Cl$_2$, DM interaction plays an important role in the gapless excitation of the ground state [36]. Here we perform further calculations on the ground state phase diagram by taking DM interaction into account. For labeling of the NNN spin sites, please refer to the Supplemental Material [26].

The ground state phase diagrams with different values of DM interaction are shown in Fig. 10. We find that the sign of DM interaction has no effect on the phase diagram. Since the DM interaction in YbOCl is almost along the *c* axis [25], the sign of the DM interaction will only change the phase of the ground state wave function but not the magnetic moments in real space. More importantly, when the DM interaction $|\vec{D}_{ij}|$ is less than 0.4 K, the phase diagram does not change much compared to the one without DM interaction. When $|\vec{D}_{ij}|$ is 0.6 K, the Col-I phase disappears and the region of the SD phase has been expanded. When $|\vec{D}_{ij}|$ continues to increase above 0.8 K, the areas of the Col-I, Col-II, and Neel phases decrease significantly, while the AFM-120° phase dominates the ground state phase diagram. Therefore, the large DM interaction causes the magnetic moments to be more inclined and eventually to lie in the *ab* plane. It is easy to understand that although the D-vector is along the *c* axis, the cross product produces the DM interaction with $S_x$ and $S_y$. It seems that the zigzag phase is less affected by DM interaction. The main reason for this may be related to the fact that the zigzag phase is located in the upper left and upper right corners of the phase diagram. Here the values of $J_{\pm\pm}$ or $J_{z\pm}$ are relatively large, so the DM interaction may be insufficient to change the zigzag phase. Not surprisingly, the areas of the SD phase are significantly reduced with increasing DM interaction, since the DM interaction behaves like an in-plane magnetic field and has a tendency to force the system into an ordered state.

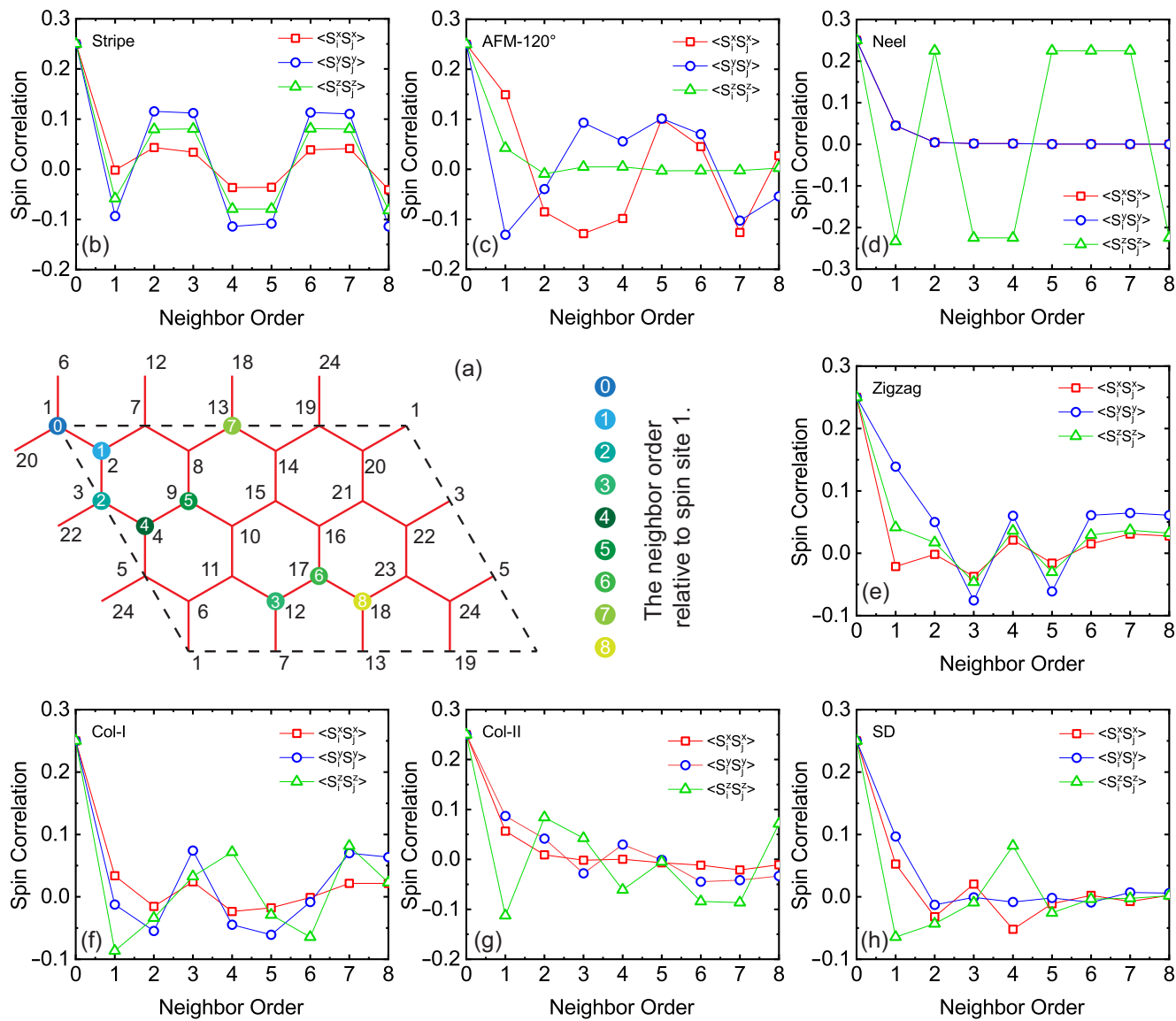

FIG. 9. (a) The colored spin sites mark the neighbor order relative to spin site No.1. (b)–(h) The evolution of spin correlation with distances or neighbor orders. The red, blue, and green lines mark spin correlations of $\langle S_i^x S_j^x \rangle$, $\langle S_i^y S_j^y \rangle$, and $\langle S_i^z S_j^z \rangle$, respectively.

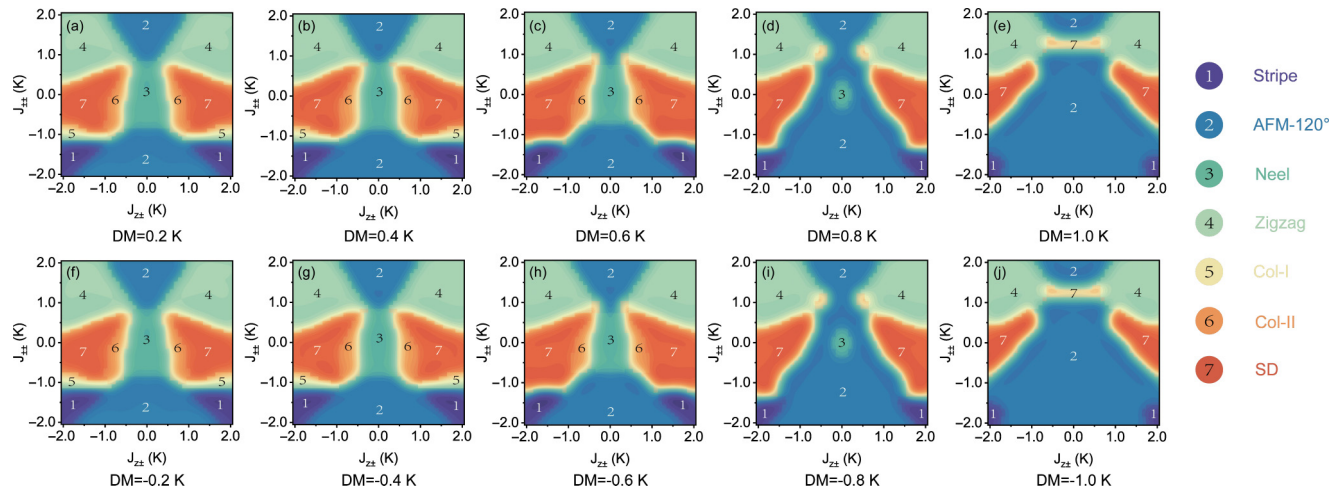

FIG. 10. Ground state phase diagram with different values of DM interaction.

## VIII. SUMMARY

In this paper, we present a clear and self-consistent understanding of the magnetism at finite temperatures for YbOCl, a Kitaev compound that was reported recently. We make the symmetry analysis on the lattice structure of YbOCl, which gives an appropriate anisotropic low-energy effective spin Hamiltonian. We measure the magnetizations at finite temperatures, including M/H-T data and M-H data. With the MF approximation and the 12-site FD method, we are able to simulate and cross check the thermodynamics data and determine the two spin-exchange interactions $J_\pm \sim -0.3$ K and $J_{zz} \sim 1.6$ K and the Landé factors $g_{ab} \sim 3.4$ and $g_c \sim 2.9$. We further perform high magnetic field ESR measurements. We develop the MF scenario and derive the MF self-consistent equations. This brings us a complete and self-consistent understanding of the ESR experiments.

Based on the experimental studies, we employ 24-site ED to calculate the ground state phase diagram in terms of $J_{\pm\pm}$ and $J_{z\pm}$. A rich ground state phase diagram containing seven magnetic phases is revealed. The magnetic phases include both the magnetically ordered phases and the disordered one. Among them, the spin-disordered phase may be related to the quantum spin liquid phase. We also investigate the influence of DM interaction on the phase diagram. The calculations show that the phase diagram with a small DM interaction does not change much compared to the phase diagram without DM interaction. When the DM interaction is large, the AFM-120° phase dominates the phase diagram.

## ACKNOWLEDGMENTS

This work was supported by the National Key Research and Development Program of China (Grant No. 2017YFA0302904), the National Science Foundation of China (Grants No. U1932215 and No. 11774419), and the Strategic Priority Research Program of the Chinese Academy of Sciences (Grant No. XDB33010100). Q.M.Z. acknowledges the support from Users with Excellence Program of Hefei Science Center and High Magnetic Field Facility, CAS. The numerical packages for classical mean-field and self-consistent equations calculations, 12-site FD calculations, and 24-sites ED calculations, are designed and developed based on MathWorks MATLAB software [37] (Academic License for Renmin University of China).